\newif\ifjcs
\documentclass[11pt]{article}
\pdfoutput=1

\usepackage{graphicx}
\usepackage{url}
\usepackage{siunitx}
\usepackage{amssymb}
\usepackage{xcolor}
\usepackage{lipsum}
\usepackage{comment}

\ifjcs
  \journal{J Computational Science}
\else
  \usepackage[margin=1in]{geometry}
  \usepackage{authblk}

  \title{\textbf{Translating the Grid:\\How a Translational Approach Shaped the Development of Grid Computing}}
  \author[1,2]{Ian Foster}
  \author[3,4]{Carl Kesselman}
  \affil[1]{Data Science and Learning Division, Argonne National Lab, Lemont, IL 60439, USA}
  \affil[2]{Department of Computer Science, University of Chicago, Chicago IL 60637, USA}
  \affil[3]{Information Sciences Institute, Marina del Rey, CA 90292, USA}
  \affil[4]{Department of Industrial and Systems Engineering, U.\ Southern California, Los Angeles, CA 90089, USA}
  \date{}
\fi

\begin{document}

\ifjcs
  \begin{frontmatter}
\else
  \maketitle
\fi


\begin{abstract}
A growing gap between progress in biological knowledge and improved health outcomes  inspired the new discipline of translational medicine, in which the application of new knowledge is an explicit part of a research plan. Abramson and Parashar argue that a similar gap between complex computational technologies and ever-more-challenging applications demands an analogous discipline of translational computer science, in which the deliberate movement of research results into large-scale practice becomes a central research focus rather than an afterthought.  We revisit from this perspective the development and application of grid computing from the mid-1990s onwards, and find that a translational framing is useful for understanding the technology's development and impact.  We discuss how the development of grid computing infrastructure, and the Globus Toolkit, in particular, benefited from a translational approach. We identify lessons learned that can be applied to other translational computer science initiatives.
\end{abstract}

\ifjcs

  \end{frontmatter}
\fi


\section{Introduction}\label{sec:intro}

Abramson and Parashar (A\&P) argue in a recent article~\cite{abramson2019translational} for the formalization of translational computer science (TCS) as a complement to traditional modes of computer science research. 
Translation as a concept emerged in medicine in the 1990s~\cite{minna1996translational}.
Observing a growing number of biomedical research discoveries that had not been exploited to improve clinical outcomes~\cite{pober2001obstacles}, 
biomedical researchers argued for new research approaches in which the application
of research results is an explicit part of the research plan---that indeed, have the goal of translating research results into ``new approaches for prevention, diagnosis, and treatment of disease''~\cite{woolf2008meaning}.
Translational medicine has since become a major part of the biomedical research enterprise.
A\&P argue that in computer science also,
the scale of technological disruptions and the increasing centrality of
computing and data in society, ``[make] it desirable that we closely couple cycles of innovation between computer science and other disciplines, with the potential to significantly accelerate the transformative impact of computer science.''
A\&P discuss how effective translation requires a thoughtful and deliberate
staging of technologies from the \emph{laboratory} to the \emph{locale}---TCS analogs of translational medicine's
\emph{bench} and \emph{bedside}.

In this article, we revisit, from the perspective of translational computer science, our work from the mid-1990s onwards in grid computing. 
We first summarize the problems that we sought to solve and
the approach that we took to solving those problems.
Then we examine this work from a TCS perspective, 
noting characteristics that seem to have correlated with success in grid computing projects, and discussing issues such as publication, student engagement, and sustainability.
We conclude with some personal reflections on translational computer science.

\section{Context: Grid Computing and Virtual Organizations}
\label{sec:grid}

The work that we discuss in this article had its origins in the early 1990s,
when close-to-gigabit/second (Gbps) networks started to become available to the research community.
Initially, these networks were used primarily for network protocol research, but as
often in computing, order-of-magnitude changes in technology drove new thinking about what could be done and how it might best be accomplished~\cite{catlett1992search,kleinrock1992latency}.

For us, these developments presented an opportunity to explore new approaches to the question of how
scientific collaborations might function without regard for distance or institutional boundaries.
When expressed in such general terms, this problem is far from new, having motivated, for example,
the development of the ARPANET in the 1960s for
sharing access to computers~\cite{roberts1988arpanet}.
However, with the emergence of gigabit networks, it took on a new flavor.
It became feasible to 
think of research as occurring in a distributed laboratory.
This new view of research environments
both spurred consideration of, and was motivated by, increasingly team-based research practices~\cite{national1993national,Cummings2008}.

\begin{figure}
\centering
\includegraphics[width=\textwidth,trim=5mm 110mm 16mm 35mm,clip]{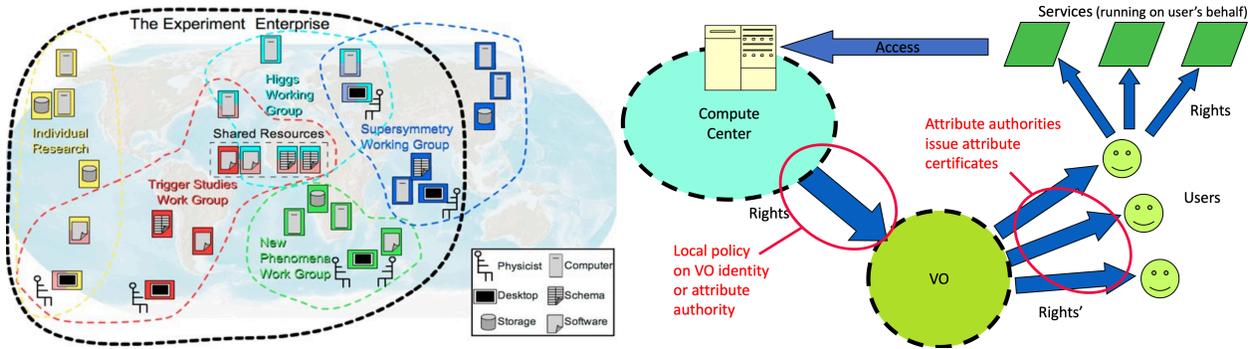}
\caption{Virtual organization concepts in Grid computing.
Left: Different virtual teams within a physics collaboration; each can access different subsets of available resources. (Credit: Harvey Newman.) 
Right: Realization via protocols and software of rights delegation within a virtual organization (VO). A resource provider grants certain rights to a VO,
which issues attribute certificates to its members; those members create processes that use those certificates to access the resource.\label{fig:vos}}
\end{figure}

In response to these transitions, 
we developed what came to be known as grid computing~\cite{anatomy}.
We approached this problem as computer scientists, looking at system architectures, security protocols, and distributed systems concepts. 
We examined how the Grid could support the creation of \emph{virtual organizations}~\cite{Cummings2008}---distributed teams of researchers that collaborated by sharing network-connected computational resources, data, and services: see Figure~\ref{fig:vos}. (This term, we later discovered, had already been coined in the management literature~\cite{ahuja1999network}.)
As with many computer science research projects, we explored these issues by creating software that implemented our ideas. 
The resulting Globus Toolkit~\cite{anatomy}
addressed needs such as secure and reliable dispatch of computational tasks to, and movement of data from and to, remote computers;
authentication and authorization of people, and for programs acting on behalf of people, within dynamic, cross-institutional environments; and
resource discovery within these environments. 
However, because of the core focus on enabling team-based research and discovery, even the earliest work in Grid computing, and in particular our development of Globus, did not stop with the development of tools, but rather actively engaged in evaluation of the tools being used to solve in use.

\section{Grid Computing as Translation}

We discuss the this work in grid computing from the perspective of translational computer science,
comparing and contrasting it with work in applied and
interdisciplinary computer science,
systems-level science, and translational medicine.

\subsection{Different questions: Applied vs.\ Translational Research}

A\&P compare and contrast applied and translational research. In the former, a research problem has a ``\emph{potential} real-world application (driver)'' (emphasis added); in the latter, the ``findings are applied as a specific phase of the research plan.'' This subtle distinction was of great importance in grid computing.

High-speed networks spurred much exploratory work in which researchers investigated specific technical problems 
(e.g., the best network protocols for high-speed transport) and 
new types of applications (e.g., distributed simulation, 
telemedicine, collaboratories).
Other researchers explored new distributed computing architectures~\cite{smarr1992metacomputing}.
They sought to ask questions like:
How can I move data at 1 Gb/s over a wide area network? 
Can I decompose a climate model into components that run in different cities?
How can people in different locations manipulate a scientific simulation from within a virtual reality environment~\cite{defanti1996overview}?
As part of this work, researchers built many inventive one-off solutions to distributed computing problems.
However, because this work was applied, not translational, 
little was learned about how the new methods worked in practice.

\begin{figure}
\centering
\includegraphics[width=\textwidth,trim=42mm 143mm 67mm 31mm,clip]{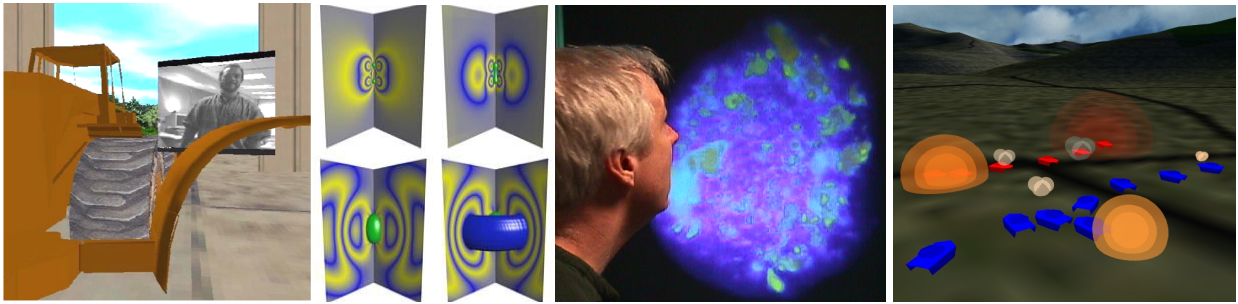}
\caption{Grid application examples, circa 2000. From left to right: tele-immersion for distance collaboration; transatlantic remote visualization of numerical relativity calculation; online analysis of instrument data; record-setting distributed simulation linking 1352 processors over 13 supercomputers at nine sites to simulate over 100,000 virtual entities.\label{fig:exs}}
\end{figure}

We asked different questions and as a result, obtained different answers. 
In particular, observing how simple Internet protocols had unleashed great creativity in networked applications, we asked: what similarly simple mechanisms might enable a comparable explosion in distributed collaboration applications, including but not limited to those illustrated in Figure~\ref{fig:exs}?  
We approached this central question from the perspective of a \textit{sociotechnical system}~\cite{sociotechnical}, considering not only technical issues
(e.g., is our solution performant, reliable, and secure?)\ but also human and community  factors (e.g., will application developers and users adopt it? Will resource providers deploy it?).
Realizing that evaluation at scale was essential to answering these questions,
we decided from the beginning that we would have to work directly with domain scientists,
application developers, and resource providers on deployment, application, and evaluation. 
In other words, we organized our research as translational computer science.

Our research over more than a decade involved work with several different communities, each with its specific locale (in the language of A\&P), unique characteristics, and culture. 
For example, the Grid Physics Network (GryPnyN)~\cite{avery2001griphyn}, 
National Virtual Observatory (NVO)~\cite{szalay2001national}, Southern California Earthquake Center (SCEC)~\cite{SCEC}, and Earth System Grid~\cite{williams2009earth}
addressed distributed computing challenges in the contexts of (primarily) high energy physics, astronomy, seismology, and climate science, respectively. 
In each project, 
we addressed computer science questions in collaboration with a community with its own high-level goals and objectives, such as creating a global data sharing and computing platform that was used to discover the Higgs Boson, or sharing large-scale simulation results to increase understanding of climate change.

We see this joint pursuit of a common goal as a key characteristic
that differentiates translational computer science from foundational and applied computer science.
This difference is apparent, for example, in the contrasting trajectories of peer-to-peer and grid computing, two research thrusts that coexisted during much of the 2000s.
While these two research threads and communities can be contrasted along many dimensions, 
one concerns the primary motivation:
to generate new knowledge in computer science 
or to produce new methods for solving important societal (primarily scientific)
problems.

Peer-to-peer computing research was concerned above all with realizing new algorithmic approaches to fault tolerance, security, and computational models~\cite{milojicic2002peer}. 
While grid computing researchers built infrastructure for evaluating research concepts in the context of operational scientific applications~\cite{johnston1999grids, gagliardi2002european, catlett2008teragrid}, peer-to-peer computing built large-scale experimental infrastructure~\cite{chun2003planetlab}. 
P2P research contributions such as distributed hash tables ultimately had a significant impact on operational large-scale systems~\cite{chang2008bigtable, sivasubramanian2012amazon}, but 
initial P2P research was not motivated by such goals, with the result that technology transfer took longer and could not inform the computer science research.

Grid computing, in contrast, created infrastructure that allowed thousands of scientists to conduct world-class science.
It also motivated computer science contributions, such as
the computational grid architecture, 
with its mechanisms for secure and reliable remote computation;
virtual organization constructs for establishing and managing the fluid and distributed trust relationships that characterize scientific collaborations;
mechanisms for performant and reliable data transfer over increasingly fast networks; and
new approaches to specifying and executing workflows~\cite{foster2010history}.

\subsection{System-level Science}

One reason for the emergence of grid computing and related technologies is the growing importance of \emph{system-level science}: science that requires the integration of diverse sources of knowledge about the constituent parts of a complex system in order to  understand  the system’s properties as a whole.
SCEC, cited above, is an example par excellence of such integration,
but many other problems also have that property, 
from understanding climate change to engineering active matter. 

Writing about the growing importance of system-level science in 2008~\cite{foster2006scaling}, 
we noted that its success depended increasingly on the availability of enabling
infrastructure that could permit system-level scientists to ask and answer questions
with required agility. 
We quoted Robert Calderbank (personal communication): 
``Sometimes through heroism, you can make something work. However, understanding why it worked, abstracting it, making it a primitive is the key to getting to the next order of magnitude of scale.''
We argued for new approaches to scientific computing infrastructure to enable 
scaling of system-level science in terms of the scope of the problems addressed, the number of resources engaged, the number of participants, and the breadth of its application.
The creation of this new infrastructure required translational computer science.

\subsection{Commitment and Sustainability}

An important element of the translation process is a 
commitment by computer scientists that they will be there for the long haul.
Software must be suitable for large-scale use and must be sustained over time.
As the development, maintenance, and continued enhancement of software are all expensive, these needs can be challenging to meet. 
As a consequence, our work in grid computing also involved experiments and innovations in the software sustainability. 
For example, we were early advocates for open source, 
at a time when 
open source was often viewed with suspicion in the licensing offices of many research institutions.
Believing in inter-institutional and international collaboration, we
established the Globus Alliance for grid software development.
Later, we established a company, Univa, to provide commercial support for grid software.
From 2010 onwards, we
established the Globus service as a subscription-supported, not-for-profit, university-based service~\cite{chard2016globus}; so far, that latter approach has been the most successful~\cite{foster2013software}. 
The point here is not that the Globus approach is right for everyone,
but that sustainability concerns must be considered carefully in TCS.

\subsection{Translational vs.\ Interdisciplinary Research}

We discussed the difference between applied and translational computer science above. 
In trying to understand the role of TCS, it is also natural to ask how it may differ from interdisciplinary research. 
The National Academies define interdisciplinary research as 
\begin{quote}
a mode of research by teams or individuals that integrates information, data, techniques, tools, perspectives, concepts, and/or theories from two or more disciplines or bodies of specialized knowledge to advance fundamental understanding or to solve problems whose solutions are beyond the scope of a single discipline or area of research practice~\cite{engineering2004facilitating}.
\end{quote}

Note that inter-disciplinary research need not be translational. 
True TCS means that the desired research result must require advances in basic computer science knowledge, not just the application of existing knowledge. As such, it can require a careful balancing act, in that computer science advances must be timed and sequenced so as to meet the timelines and constraints of the community being translated to. We discuss sequencing issues below. 

In general, we have observed two key requirements for effective TCS.
First, computer science research results must be generated in a manner that is consistent with the requirements of the locale to which the translation is being conducted, and translation must produce a usable research artifact with a direct perceived and actual benefit to the receiving community. 
Second, the receiving community must be a willing participant in the process. 
They must be prepared to expend effort and perhaps sacrifice some short-term efficiency for long-term gain, which---a further complicating factor---may benefit others beyond themselves.  
Note that these considerations are much those that accompany like translation in a clinical setting. Participating in a clinical trial always involves both explicitly defined risk and intended benefits to society, both of which are typically enumerated in the consent form associated with the trial.

\subsection{Translational Computer Science is Different}

TCS differs from translational medicine in ways other than a lack of white coats.
In the medical literature, translational research is often decomposed into four concrete steps, named imaginatively as T0 through T4~\cite{T0to4},
and progressing from basic research to translation to communities.
These steps are generally applied in sequence. 
A major factor in this sequencing is the legal, ethical, and regulatory requirements associated with human subject research, which at each step require well-defined studies that have been pre-approved by an institutional review board. 

In contrast, software development frequently employs an agile methodology~\cite{martin2002agile}. 
Our experience is that translation is usually best performed as an iterative process, and that unlike biomedical applications with distinct states, the methods of iterative, agile development are powerful mechanisms of translation. 
Each iteration provides a valuable opportunity for feedback of knowledge that can drive the direction and results of the foundational research. For example, in Grid computing, much work was done on job scheduling.  Traditionally, this research would have been evaluated using synthetic or representative workloads. In the translational setting, this research can be evaluated \emph{in situ}. We learned from such studies that the scheduling problem that we thought needed to be solved never occurred in practice, while scenarios initially dismissed as corner cases were in fact important.

Some communities are not comfortable with the idea of agile development, and the effective translational researcher must proceed accordingly. 
For example, in NEES (discussed further below), the structural and civil engineering community were fundamentally uncomfortable with the idea of iterative improvement~\cite{finholt2006if,spencer200818}.

\section{Impacts and Lessons Learned}

We speak briefly to impacts and note some lessons learned.

\subsection{Measuring Impacts}

Evaluating the impact of research is difficult,
and perhaps especially so in computer science, due to the often rapid evolution of technology and the frequent re-emergence of old ideas in new forms.
For example, utility computing~\cite{douglas1966challenge}, 
grid computing, and cloud computing~\cite{mell2011nist}, first described in the 1960s,
1990s, and 2000s, respectively, address similar intellectual themes
but innovate in different ways.

Nevertheless, we can point to several metrics that suggest that the Grid research described here has delivered both useful computer science knowledge and translational benefits to the scientific domains with which we engaged.
First, in computer science, 
the number of citations to relevant scientific articles suggest that
the work reviewed here had a great deal of impact, 
with \num{10000}s of citations in many thousands of scientific articles.

\begin{figure}
\centering
\includegraphics[width=\textwidth,trim=33mm 138mm 88.2mm 38mm,clip]{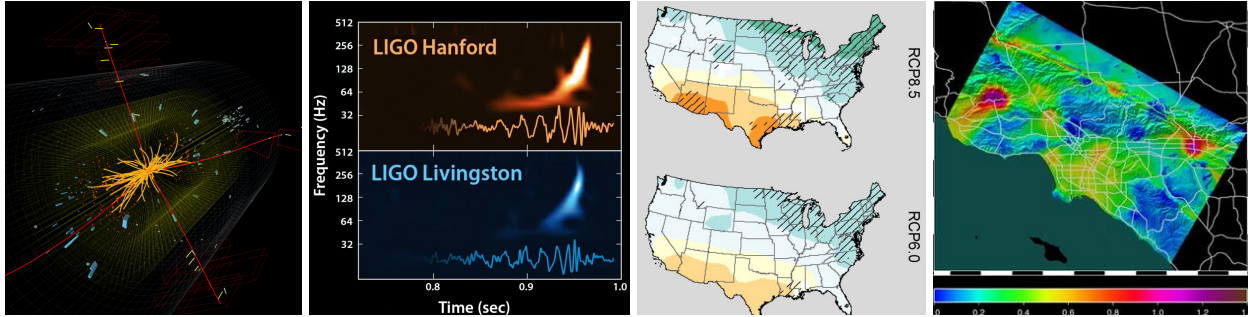}
\caption{Examples of Grid impact on science. 
From left to right: Collision event at LHC, signifying Higgs boson discovery announced July 4, 2012;
data at two LIGO sites on Sept 14, 2015, indicating detection of gravitational wave signal GW150914~\cite{abbott2016observation};
climate science based on ESG data, showing warming projections for the US under different scenarios;
hazard map for the Los Angeles region, computed with CyberShake~\cite{graves2011cybershake}.\label{fig:success}}
\end{figure}

From the perspective of the target domains,
scientific contributions that have been enabled, in part, by these developments
include the discovery of the Higgs boson, aided by the Worldwide LHC Computing  Grid~\cite{aad2012observation,chatrchyan2012observation};
the detection of gravitational waves, aided by the LIGO Data Grid~\cite{abbott2016observation};
improved understanding of climate change, aided by the Earth System Grid Federation~\cite{williams2009earth}; and better estimates of earthquake hazards~\cite{graves2011cybershake}: see Figure~\ref{fig:success}.

We can also point to impacts on scientific infrastructure more broadly,
via for example the US National Science Foundation's TeraGrid~\cite{catlett2008teragrid} and XSEDE networks~\cite{towns2014xsede}, and 
the widespread deployment of data transfer nodes that leverage grid technologies
for high-speed data transfer~\cite{dart2014science}.

\subsection{Locale as Playground and Straitjacket}

As noted earlier, TCS implies a need for a locale into which research results are to be translated.
It is this engagement with a specific locale that enables both the rapid feedback and technology transfer that are essential to effective TCS,
by permitting researchers to (in A\&P's words) ``test their ideas at scale with real-world constraints.''

However, the focus on a specific locale can also constrain computer science research in ways that hinder its development by imposing too many constraints.
For example, in the case of grid computing,
the state of scientific computing infrastructure was such that supercomputer centers became perforce the primary locale for early translation work.
Such centers had then, and continue to have today, positive properties, such as substantial computing resources, talented and engaged staff,
and co-location with leading-edge application teams. However, other factors, such as specialized computing environments, oversubscribed resources, and constrained economic models made experiments with on-demand computing difficult.
Thus it is not surprising that large-scale deployments of on-demand computing ultimately occurred outside science, in the form of commercial cloud computing based on low-cost commodity hardware and pay-for-use economic models.

\subsection{How Translation Can Fail}

TCS benefits from a tight coupling of intended research result and target  translation locale. When this coupling is lacking, translation can fail. 

We observed this result in the NEESGrid project~\cite{spencer200818}, in which we attempted to engage in TCS with a broad community of earthquake engineers. Rather than focusing on translation into a specific locale and then translate to a community, the structure of this project required that we simultaneously both develop computer science concepts (such as integrating computational modeling with on-line physical testing facilities) and translate to multiple significantly different sub-communities (structural engineering, geotechnical engineering, test facility developers). 
Much of the work took place with only a superficial engagement of the members of the targeted locales. 
Although valuable computer science concepts were developed, the translational part was extremely challenging and ultimately failed. 
Subsequent analysis showed that these issue resulted from a misalignment of incentives, expectations, and communications~\cite{finholt2006if}. 
The lessons of NEESGrid are clearly that understanding the social aspects of translation are critical and ignoring these elements can result in failure to translate. 

\subsection{Team Structure; Engaging Students}

Effective TCS requires
substantial software engineering talent.
This need introduces numerous challenges, from raising the funds to pay for (expensive) talent to recruiting, retaining, and compensating those individuals and providing them with an effective career path.
We have been fortunate to have been supported by funding agencies and to work in institutions that have been supportive of these needs.

We have also struggled with how best to engage students in translational research.
The truly important questions often become clear only over time as translation proceeds.
Some questions that seemed important initially turn out not to be essential to effective translation (and thus, in the interest of efficiency, should be dropped), while other apparently trivial questions turn out to be fundamental, and yet others 
turn out to be straightforward (but essential) engineering.
While these characteristics of TCS can make it rewarding, 
they may not align with a student's need to pursue a single problem deeply. 

We have found it effective to organize teams in which research scientists and research programmers undertake much of the translation process,
while Ph.D.\ students tackle questions that are exposed by translational work but are not essential to ultimate success. 
Students then benefit from the data and expertise provided by their connection to the translational project, without being constrained by a need to deliver results on a specific timetable.
Two dangers to which we must be alert are, on the one hand, the student becoming pulled into engineering and, on the other, the student becoming disconnected from the translational research team. 
Different considerations can apply in the case of MS students,
for which tackling a substantial engineering problem in a friendly and supportive environment can be an excellent learning experience.

\section{Conclusions}

Research was long categorized as either \emph{pure} or \emph{applied}~\cite{morse1950pure,feibleman1961pure}.
Donald Stokes' influential analysis of scientific practice 
introduced a two-dimensional 
classification 
according to whether or not research seeks fundamental understanding,
and whether it is motivated by curiosity or by a desire to benefit society~\cite{stokes2011pasteur}.
We may view the concept of translational science to be a further refinement of Stokes' analysis, according to which the question of how well new knowledge works in practice becomes a vital metric when evaluating research success.  

Translation has become an established fixture in medicine.
In the U.S., the National Institutes of Health operates a National Center for Advancing Translational Sciences, and translational medicine institutes exist in many universities and medical centers.
The idea of translation as a distinct activity, worthy of study in its own right,
has not yet been impactful in other domains, but A\&P make a strong case that it should.
TCS is not yet a discipline,
but as it develops, it will be instructive to compare and contrast its methods and outcomes with those of translational medicine. 
One difference that we think is likely important is the opportunities offered by computational technologies for rapid iteration among different stages in the translation pipeline.

We conclude with a personal note. 
We have on occasion seen our work dismissed as mere engineering.
Nevertheless, we have persisted in our belief that it is by building large systems that we both identify the real problems and deliver benefits to society. 
Thus, like Moliere's Jourdain, who is delighted to discover that ``[f]or more than forty years I have been speaking prose while knowing nothing of it''~\cite{Moliere},
we are delighted to discover that over the last 30 years, 
we have not been mucking around with computers, but practicing translational computer science.

\section*{Acknowledgements}

We dedicate this article to the memory of our talented friend and long-time partner in translational computer science, the late Steven J.\ Tuecke.
The development of this article was supported in part by funding from 
the US Department of Energy under contract DE-AC02-06CH11357.

\ifjcs
\else
    \bibliographystyle{plain}
\fi
\bibliography{refs.bib}

\begin{thebibliography}{10}

\bibitem{SCEC}
{Southern California Earthquake Center}, 2020.
\newblock \url{http://scec.org}.

\bibitem{T0to4}
What are the {T0 to T4} research classifications?, 2020.
\newblock
  \url{https://ictr.wisc.edu/what-are-the-t0-to-t4-research-classifications}.

\bibitem{abramson2019translational}
David Abramson and Manish Parashar.
\newblock Translational research in computer science.
\newblock {\em Computer}, 52(9):16--23, 2019.

\bibitem{ahuja1999network}
Manju~K Ahuja and Kathleen~M Carley.
\newblock Network structure in virtual organizations.
\newblock {\em Organization Science}, 10(6):741--757, 1999.

\bibitem{aad2012observation}
{ATLAS Collaboration}.
\newblock Observation of a new particle in the search for the standard model
  {H}iggs boson with the {ATLAS} detector at the {LHC}.
\newblock {\em Physics Letters B}, 716(1):1--29, 2012.

\bibitem{avery2001griphyn}
Paul Avery and Ian Foster.
\newblock The {GriPhyN} project: Towards petascale virtual data grids.
\newblock Technical report, Technical Report GriPhyN-2001-15, 2001.

\bibitem{sociotechnical}
Gordon Baxter and Ian Sommerville.
\newblock {Socio-technical systems: From design methods to systems
  engineering}.
\newblock {\em Interacting with Computers}, 23(1):4--17, 08 2010.

\bibitem{catlett1992search}
Charles~E Catlett.
\newblock In search of gigabit applications.
\newblock {\em IEEE Communications Magazine}, 30(4):42--51, 1992.

\bibitem{catlett2008teragrid}
Charlie Catlett, William~E Allcock, Phil Andrews, Ruth Aydt, Ray Bair, Natasha
  Balac, Bryan Banister, Trish Barker, Mark Bartelt, Pete Beckman, et~al.
\newblock Tera{G}rid: Analysis of organization, system architecture, and
  middleware enabling new types of applications.
\newblock In {\em High Performance Computing and Grids in Action}, pages
  225--249. IOS press, 2008.

\bibitem{chang2008bigtable}
Fay Chang, Jeffrey Dean, Sanjay Ghemawat, Wilson~C Hsieh, Deborah~A Wallach,
  Mike Burrows, Tushar Chandra, Andrew Fikes, and Robert~E Gruber.
\newblock Big{T}able: A distributed storage system for structured data.
\newblock {\em ACM Transactions on Computer Systems}, 26(2):1--26, 2008.

\bibitem{chard2016globus}
Kyle Chard, Steven Tuecke, and Ian Foster.
\newblock Globus: Recent enhancements and future plans.
\newblock In {\em XSEDE16 Conference}, pages 1--8, 2016.

\bibitem{chun2003planetlab}
Brent Chun, David Culler, Timothy Roscoe, Andy Bavier, Larry Peterson, Mike
  Wawrzoniak, and Mic Bowman.
\newblock Planet{L}ab: An overlay testbed for broad-coverage services.
\newblock {\em ACM SIGCOMM Computer Communication Review}, 33(3):3--12, 2003.

\bibitem{chatrchyan2012observation}
{CMS Collaboration}.
\newblock Observation of a new boson at a mass of 125 {GeV} with the {CMS}
  experiment at the {LHC}.
\newblock {\em Physics Letters B}, 716(1):30--61, 2012.

\bibitem{Cummings2008}
Jonathan Cummings, Thomas Finholt, Ian Foster, Carl Kesselman, and Katherine
  Lawrence.
\newblock {\em Beyond Being There: A Blueprint for Advancing the Design,
  Development, and Evaluation of Virtual Organizations}.
\newblock 2008.

\bibitem{dart2014science}
Eli Dart, Lauren Rotman, Brian Tierney, Mary Hester, and Jason Zurawski.
\newblock The {Science DMZ}: A network design pattern for data-intensive
  science.
\newblock {\em Scientific Programming}, 22(2):173--185, 2014.

\bibitem{defanti1996overview}
Thomas~A DeFanti, Ian Foster, Michael~E Papka, Rick Stevens, and Tim Kuhfuss.
\newblock Overview of the {I-WAY}: Wide-area visual supercomputing.
\newblock {\em International Journal of Supercomputer Applications and High
  Performance Computing}, 10(2-3):123--131, 1996.

\bibitem{feibleman1961pure}
James~K Feibleman.
\newblock Pure science, applied science, technology, engineering: An attempt at
  definitions.
\newblock {\em Technology and Culture}, 2(4):305--317, 1961.

\bibitem{finholt2006if}
Thomas~A Finholt and Jeremy~P Birnholtz.
\newblock If we build it, will they come? {T}he cultural challenges of
  cyberinfrastructure development.
\newblock In {\em Managing Nano-bio-info-cogno Innovations}, pages 89--101.
  Springer, 2006.

\bibitem{foster2006scaling}
Ian Foster and Carl Kesselman.
\newblock Scaling system-level science: Scientific exploration and {IT}
  implications.
\newblock {\em Computer}, 39(11):31--39, 2006.

\bibitem{foster2010history}
Ian Foster and Carl Kesselman.
\newblock The history of the grid, 2010.
\newblock \url{https://pdfs.semanticscholar.org/6a92}.

\bibitem{anatomy}
Ian Foster, Carl Kesselman, and Steven Tuecke.
\newblock The anatomy of the grid: Enabling scalable virtual organizations.
\newblock {\em International Journal of High Performance Computing
  Applications}, 15(3):200--222, 2001.

\bibitem{foster2013software}
Ian Foster, Vas Vasiliadis, and Steven Tuecke.
\newblock Software-as-a-service as a path to software sustainability, 2013.
\newblock \url{https://doi.org/10.6084/m9.figshare.791604.v1}.

\bibitem{gagliardi2002european}
Fabrizio Gagliardi, Bob Jones, Mario Reale, and Stephen Burke.
\newblock European {DataGrid} project: Experiences of deploying a large scale
  testbed for e-science applications.
\newblock In {\em IFIP International Symposium on Computer Performance
  Modeling, Measurement and Evaluation}, pages 480--499. Springer, 2002.

\bibitem{graves2011cybershake}
Robert Graves, Thomas~H Jordan, Scott Callaghan, Ewa Deelman, Edward Field,
  Gideon Juve, Carl Kesselman, Philip Maechling, Gaurang Mehta, Kevin Milner,
  David Okaya, Patrick Small, and Karan Vahi.
\newblock {CyberShake}: A physics-based seismic hazard model for southern
  {C}alifornia.
\newblock {\em Pure and Applied Geophysics}, 168(3-4):367--381, 2011.

\bibitem{johnston1999grids}
William~E Johnston, Dennis Gannon, and Bill Nitzberg.
\newblock Grids as production computing environments: The engineering aspects
  of {NASA's Information Power Grid}.
\newblock In {\em 8th International Symposium on High Performance Distributed
  Computing}, pages 197--204. IEEE, 1999.

\bibitem{kleinrock1992latency}
Leonard Kleinrock.
\newblock The latency/bandwidth tradeoff in gigabit networks.
\newblock {\em IEEE Communications Magazine}, 30(4):36--40, 1992.

\bibitem{abbott2016observation}
{LIGO Scientific Collaboration and Virgo Collaboration}.
\newblock Observation of gravitational waves from a binary black hole merger.
\newblock {\em Physical Review Letters}, 116(6):061102, 2016.

\bibitem{martin2002agile}
Robert~C Martin.
\newblock {\em Agile Software Development: Principles, Patterns, and
  Practices}.
\newblock Prentice Hall, 2002.

\bibitem{mell2011nist}
Peter Mell and Tim Grance.
\newblock The {NIST} definition of cloud computing.
\newblock Technical Report SP 800-145, Computer Security Division, Information
  Technology Laboratory, National Institute for Science and Technology, 2011.

\bibitem{milojicic2002peer}
Dejan~S Milojicic, Vana Kalogeraki, Rajan Lukose, Kiran Nagaraja, Jim Pruyne,
  Bruno Richard, Sami Rollins, and Zhichen Xu.
\newblock Peer-to-peer computing.
\newblock Technical Report HPL-2002-57, HP Labs, 2002.

\bibitem{minna1996translational}
Jhon~D Minna and Adi~F Gazdar.
\newblock Translational research comes of age.
\newblock {\em Nature Medicine}, 2(9):974--975, 1996.

\bibitem{Moliere}
Moli\`{e}re.
\newblock {\em Le Bourgeois Gentilhomme}.
\newblock \'{e}dition Barbin, 1670.

\bibitem{morse1950pure}
Philip~M Morse.
\newblock Pure and applied research.
\newblock {\em American Scientist}, 38(2):253--259, 1950.

\bibitem{engineering2004facilitating}
{National Academy of Sciences, National Academy of Engineering, and Institute
  of Medicine}.
\newblock {\em Facilitating Interdisciplinary Research}.
\newblock National Academies Press, 2005.
\newblock \url{https://doi.org/10.17226/11153}.

\bibitem{national1993national}
{National Research Council}.
\newblock {\em National collaboratories: Applying information technology for
  scientific research}.
\newblock National Academies Press, 1993.

\bibitem{douglas1966challenge}
Douglas~F Parkhill.
\newblock {\em The Challenge of the Computer Utility}.
\newblock Addison-Wesley Publishing Company, 1966.

\bibitem{pober2001obstacles}
Jordan~S Pober, Crystal~S Neuhauser, and Jeremy~M Pober.
\newblock Obstacles facing translational research in academic medical centers.
\newblock {\em The FASEB Journal}, 15(13):2303--2313, 2001.

\bibitem{roberts1988arpanet}
Larry Roberts.
\newblock The {ARPANET} and computer networks.
\newblock In {\em A History of Personal Workstations}, pages 141--172. 1988.

\bibitem{sivasubramanian2012amazon}
Swaminathan Sivasubramanian.
\newblock Amazon {DynamoDB}: A seamlessly scalable non-relational database
  service.
\newblock In {\em ACM SIGMOD International Conference on Management of Data},
  pages 729--730, 2012.

\bibitem{smarr1992metacomputing}
Larry Smarr and Charles~E Catlett.
\newblock Metacomputing.
\newblock {\em Communications of the ACM}, 35(6):44--52, 1992.

\bibitem{spencer200818}
BF~Spencer~Jr, Randal Butler, Kathleen Ricker, Doru Marcusiu, Thomas~A Finholt,
  Ian Foster, Carl Kesselman, and Jeremy~P Birnholtz.
\newblock {NEESgrid}: Lessons learned for future cyberinfrastructure
  development.
\newblock In {\em Scientific Collaboration on the Internet}, page 331. MIT
  Press, 2008.

\bibitem{stokes2011pasteur}
Donald~E Stokes.
\newblock {\em Pasteur's Quadrant: Basic Science and Technological Innovation}.
\newblock Brookings Institution Press, 2011.

\bibitem{szalay2001national}
Alexander~S Szalay.
\newblock The {National Virtual Observatory}.
\newblock In {\em Astronomical Data Analysis Software and Systems X}, volume
  238, page~3, 2001.

\bibitem{towns2014xsede}
John Towns, Timothy Cockerill, Maytal Dahan, Ian Foster, Kelly Gaither, Andrew
  Grimshaw, Victor Hazlewood, Scott Lathrop, Dave Lifka, Gregory~D Peterson,
  et~al.
\newblock {XSEDE}: Accelerating scientific discovery.
\newblock {\em Computing in Science \& Engineering}, 16(5):62--74, 2014.

\bibitem{williams2009earth}
Dean~N Williams, R~Ananthakrishnan, DE~Bernholdt, S~Bharathi, D~Brown, M~Chen,
  AL~Chervenak, Luca Cinquini, R~Drach, IT~Foster, et~al.
\newblock The {Earth System Grid}: Enabling access to multimodel climate
  simulation data.
\newblock {\em Bulletin of the American Meteorological Society},
  90(2):195--206, 2009.

\bibitem{woolf2008meaning}
Steven~H Woolf.
\newblock The meaning of translational research and why it matters.
\newblock {\em JAMA}, 299(2):211--213, 2008.

\end{thebibliography}

\end{document}